# Unveiling the temporal dynamics in multi-longitudinal mode ytterbium-doped fiber lasers

**Wei Liu, Pengfei Ma, Pu Zhou and Zongfu Jiang**

College of Advanced Interdisciplinary Studies, National University of Defense Technology, Changsha, Hunan, 410073, People's Republic of China

E-mail: aiken09@163.com

**Abstract:**
In this work, we propose a unified spatio-temporal model to study the temporal dynamics in continuously pumped ytterbium-doped fiber lasers (YDFLs). Different from previously reported theories, this model is capable of obtaining the temporal evolution of an YDFL from relaxation oscillation region to relative stable region in different time scales ranging from sub-nanosecond to millisecond. It reveals that there exists dual time scale characteristics in the temporal evolution of a multi-longitudinal mode YDFL. Specifically, the temporal evolution would experience sharp change during one cavity round-trip while keep relatively stable between adjacent cavity round-trips. Representative cases are simulated to study the influences of structure parameters on the temporal dynamics and the longitudinal mode characteristics in YDFLs. Three types of temporal instabilities, i.e. sustained self-pulsing, self-mode locking, and turbulence-like pulsing, coexist in a multi-longitudinal mode YDFL. The simulation results clarify that the three temporal instabilities are all the reflectors of intrinsic characteristics of longitudinal modes superposition in multi-longitudinal mode YDFLs. In addition, the strength of the irregular sustained self-pulsing is the major issue which impacts the macroscopic temporal fluctuations in YDFLs.

## 1. Introduction

Fiber lasers provide a promising platform for the demonstration of nonlinear dynamics [1, 2]. The physics of most fiber lasers comprise a complex interaction among cyclic gain, loss, dispersion, and nonlinear effects [3]. The interplay between the dissipative and conservative effects in fiber lasers can give rise to continuous-wave (CW) or ultra-short pulsed radiation from highly stable regime to complex dynamic regime [4]. It is of both academic and experimental interest to reveal the periodic, bounded, and deterministic processes in such highly stochastic or chaotic regime, as abundant nonlinear physical behaviors could be demonstrated.

One can generally observe three kinds of temporal instabilities in a multi-longitudinal mode rare-earth doped fiber laser, i.e. sustained self-pulsing (SSP), self-mode locking (SML), and turbulence-like pulsing (TLP) [5-7]. The observation time scale provides a basis for distinguishing the three types of temporal instabilities experimentally. Generally speaking, the SSP phenomenon describes the periodic emission of laser pulses at a repetition rate associated to the relaxation oscillations [5]. The SML phenomenon is related to the occurrence of multiple longitudinal modes, which results in output pulses with a period exactly equal to one cavity round-trip time [6]. The TLP phenomenon is the superposition state of amounts of longitudinal modes with phase and amplitude fluctuations, which presents highly stochastic intensity fluctuations in sub-nanosecond time scale [7].

In previous studies, as for SSP phenomenon, it was explained through the potential mechanisms of optical pulse generation in fiber lasers, such as the ion pairs or clusters acting as saturable absorbers [8], the gain saturation effect [9], and the nonlinear effects [10]. Those works could provide a well reference for mitigating SSP effect in fiber laser systems. As for SML and TLP phenomena, the spectral-property-related theoretical analysis was applied for illustrating their occurrences [11, 12]. Overall, previous studies could explain the individual physical process of one specific temporal instability. However, different types of temporal instabilities would depend on each other in many practical fiber laser systems. In these cases, comprehensive investigation of the SSP, SML and TLP phenomena and further illustrate their internal connections may provide well reference for understanding the evolutions of multiply temporal properties and proposing effective suppression techniques.

Thus, in this work, we first aim to establish a unified spatio-temporal model to clarify the temporal dynamics in rare-earth doped fiber lasers. The model is based on the comprehensive analysis of the gain dynamics and nonlinear dynamics in fiber lasers. Take the multi-longitudinal mode ytterbium-doped fiber lasers (YDFLs) as an example, both the physical origins and properties of the three temporal instabilities are investigated numerically. We further compare the temporal dynamics and spectral properties in multi-longitudinal mode YDFLs which have different structure parameters. The comparisons clarify the internal connections among the three temporal instabilities.

## 2. Theory and numerical modeling

In conventional analysis, the evolution of the optical field could be divided into two major energy conversion processes in YDFLs: the energy absorption from the pump light to the signal light through the doped ions and the energy transformation between different spectral components during the nonlinear propagation. Generally speaking, the basic analysis of above two energy conversion processes could be fulfilled by using the time-dependent rate equations (TDREs) [13] and nonlinear propagation equations (NPEs) [14], respectively. In theory, the temporal dynamics in YDFLs could be investigated either through the temporal or spectral version, since the time-frequency properties of an optical field satisfy the basic Fourier transform relationship. However, when an YDFL is analyzed through the spectral evolution, even the temporal property in the time window of nanosecond could be commonly obtained [15], the premise of the steady-state rate equation approximation should be applied. This spectral-domain analysis could effectively deal with the fast evolution of the optical field during one cavity round-trip, while it is invalid to reflect the slow evolution of optical field among cavity round-trips. In contrast to the spectral-domain analysis, benefit from the incorporation of time-dependent active gain, the temporal-domain analysis is a robust approach for directly obtaining both fast and slow evolutions of the optical field in different time scales. In our following analysis, it will show that the refection of slow evolution of optical field is necessary for investigating the physical origins and interconnections among different types of temporal instabilities in multi-longitudinal mode YDFLs.

The formation of an YDFL could be regarded as the fact that the initial spontaneous emission noise gradually evolves and finally reaches statistically stable emission. This is a cyclic process along the cavity round-trips. The occurrences of the TLP, SML and SSP phenomena correspond to the evolutions in different time scales, i.e. during one cavity round-trip, between adjacent cavity round-trips and among an amount of cavity round-trips, respectively. Accordingly, the major task for understanding the physical origins and interconnections among different temporal dynamics is to clarify and distinguish the dual impacts of the two energy conversion processes on YDFLs in different time scales.

Based on the above considerations, by incorporating the accumulative effects of the nonlinear dynamics on the gain dynamics, the set of bidirectional spatio-temporal equations which mainly describe the evolutions of the two counter-propagation optical fields in YDFLs could be expressed as:

$$\frac{\partial N_2}{\partial t} = -\frac{N_2}{\tau} + \frac{\Gamma_p \lambda_p}{hcA_{eff}}\left(\sigma_p^a N_1 - \sigma_p^e N_2\right)P_p + \frac{\Gamma_s \lambda_s}{hcA_{eff}}\left(\sigma_s^a N_1 - \sigma_s^e N_2\right)\left(\left|A_s^+\right|^2 + \left|A_s^-\right|^2\right) \quad (1)$$

$$\frac{\partial P_p}{\partial z} + \frac{1}{v_p}\frac{\partial P_p}{\partial t} = \Gamma_p\left[\sigma_p^e N_2 - \sigma_p^a N_1\right]P_p - \alpha_p P_p \quad (2)$$

$$\pm\frac{\partial A_s^\pm}{\partial z} + \frac{1}{v_s}\frac{\partial A_s^\pm}{\partial t} + \frac{i\beta_2}{2}\frac{\partial^2 A_s^\pm}{\partial t^2} = i\gamma\left(\left|A_s^\pm\right|^2 + 2\left|A_s^\mp\right|^2\right)A_s^\pm + \frac{g_s - \alpha_s}{2}A_s^\pm + f_{sp}^\pm \quad (3)$$

where, + and − correspond to forward and backward propagations; the subscripts $p$ and $s$ stand for pump wave and signal wave, respectively; $\sigma^a$ and $\sigma^e$ are the corresponding absorption and emission cross section; $v$ is the group velocity; $N_1$ and $N_2$ are the ion densities in the ground state or excited state, $N_1+N_2=N_0$, $N_0$ is the dopant density in the fiber core; $\Gamma$ is the overlap factor; $\tau$ is the lifetime of the excited state; $\lambda$ is the wavelength, $h$ is the Planck constant, $c$ is the light velocity and $A_{eff}$ is the effective mode area of the fiber; $\beta_2$ is the second order dispersion coefficient; $\gamma$ is the nonlinear coefficient; $g$ is the gain coefficient and $\alpha$ is the loss coefficient; $f_{sp}$ is the optical field of the spontaneous emission noise.

Equations (1)-(3) describe the evolutions of the ion density in the excited state, the pump power, and two counter-propagation optical fields verse time and position in the fiber, respectively. To increase the accuracy of the simulation and investigate the practical interactions between the two counter-propagation optical fields, an iterative solution of (1)-(3) is calculated numerically using bidirectional finite difference time-domain (B-FDTD) method [16].

For a typical YDFL, the laser cavity is formed by a doped fiber of length $L$ and a pair of fiber Bragg gratings (FBGs). The dissipation of the signal light is mainly induced by FBGs, and the boundary conditions that related to the reflective spectra of FBGs are generally expressed as follows:

$$A^+(0,\omega) = \sqrt{R_1(\omega)} A^-(0,\omega) \tag{4}$$

$$A^-(L,\omega) = \sqrt{R_2(\omega)} A^+(L,\omega) \tag{5}$$

where, $A(\omega)$ is the amplitude of optical field in the frequency domain; $R_1(\omega)$ and $R_2(\omega)$ are the reflective spectra of the high reflectivity (HR) FBG and the output coupling (OC) FBG, respectively. In B-FDTD method, (4) and (5) are transformed into temporal domain through Fourier transform.

In the classical electromagnetic theory, both the amplitude and phase of $f_{sp}$ satisfy the Gaussian stochastic process. The intensity of $f_{sp}$ has zero mean value, and its variance is proportional to the gain coefficient [17]. The length of the active fiber is set to be 1 m with the dopant density of $1\times10^{26}/m^3$, and the length of the passive fiber is ignored. The reflective spectral shapes of FBGs are assumed to be Gaussian, and the bandwidths of the OC FBG and HR FBG are set to be 10 GHz and over 1 nm, respectively. Here, the bandwidth of an FBG is defined as the spectral range when the spectral intensity decreases to $e^{-1}$ of the maximum intensity. The values of major parameters used in the simulations are shown in Table I.

TABLE I

| Parameter | Value | Parameter | Value |
|---|---|---|---|
| $\lambda_p$ | 976 nm | $\lambda_s$ | 1070 nm |
| $\alpha_p$ | 0.003 m$^{-1}$ | $\alpha_s$ | 0.005 m$^{-1}$ |
| $\Gamma_p$ | 0.0064 | $\Gamma_s$ | 0.85 |
| $N$ | $1\times10^{26}/m^3$ | $\tau$ | 0.85 ms |
| $L$ | 1 m | $\gamma$ | 1.5 W$^{-1}$/km |
| $\beta_2$ | 20.4 ps$^2$/km | $v$ | $2.04\times10^8$ m/s |
| $\sigma_{as}$ | $3.93\times10^{-27}$ m$^2$ | $\sigma_{es}$ | $3.31\times10^{-25}$ m$^2$ |
| $\sigma_{ap}$ | $4.37\times10^{-24}$ m$^2$ | $\sigma_{ep}$ | $4.42\times10^{-24}$ m$^2$ |
| $R_1$ | 0.99 | $R_2$ | 0.1 |
| $dt$ | $1\times10^{-11}$ s | $T_m$ | $>4\times10^{-3}$ s |

$R_1$ and $R_2$ are the reflectivity of the HR FBG and OC FBG, respectively; $dt$ is the time resolution and $T_m$ is the time window in the simulation.

## 3. Temporal dynamics in YDFLs

For the parameter values listed in Table I, the threshold pump power is about 0.11 W. The operation state of an YDFL would become relatively stable when the pump power is over several times of the laser threshold. To illustrate the basic properties of the temporal dynamics in YDFLs, we analyze the temporal evolution of the YDFL at a pump power of 0.45 W (relatively stable state), and the influence of the pump power on the temporal dynamics of YDFLs would be analyzed in the following section. Fig. 1 shows the normalized spatio-temporal distributions of the pump power and the signal power when the YDFL works stably at a pump power of 0.45 W. The horizontal and vertical coordinates in Fig. 1 correspond to the spatial distribution along the fiber and the temporal evolution verse time, respectively. Here, the power is normalized through its maximum value. As shown in Fig. 1(a), the pump power decreases along the fiber and keeps nearly unchanged verse the evolution time coordinate. As shown in Fig. 1(b), the signal power experiences rapid and quasi-periodic change over time at each arbitrary position along the fiber. Therefore, even when the pump power in the YDFL remains stable over time, the signal power could still exhibit sharp change. In the following analysis, the temporal evolution located at the output port is selected to analyze the temporal dynamics in YDFLs.

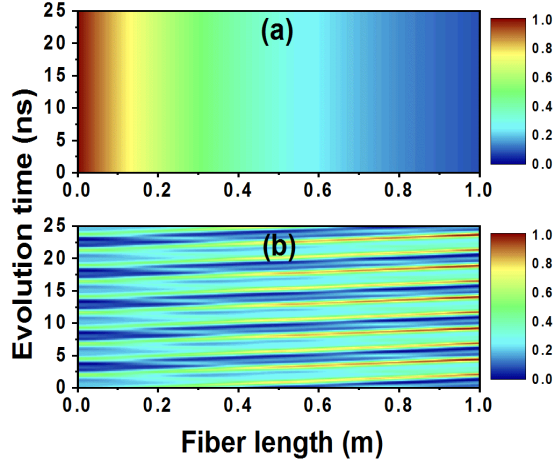

**Fig. 1.** The normalized spatio-temporal distributions of the pump power and the signal power: (a) the pump power; (b) the signal power

Fig. 2 shows the output characteristics in the YDFL at a pump power of 0.45 W with different time resolutions. The left and the right columns correspond to a time resolution of 0.1 μs and 0.01 ns, respectively. In both cases, there are three regions for the overall temporal evolutions in the YDFL: (A) the building-up region; (B) the relaxation oscillation region; (C) the relative stable region [see in Fig. 2(a) and 2(d)]. In the building-up region, the laser power is quite weak and only consists of amplified spontaneous emission noise. The relaxation oscillation region begins at about 0.24 ms after the building-up region and continues for about 1.5 ms before reaching the relative stable region. In the relative stable region, the output power keeps relatively stable. Therefore, the temporal dynamics of the YDFL in all the three regions could be studied in detail through this model. In most applications, the performance of an YDFL is mainly determined by its temporal dynamics in the relative stable region. Thus, we mainly focus on analyzing the temporal dynamics of the YDFLs in the relative stable region in the following sections.

To understand the laser behavior in more detail, we cut out a section of data from the relative stable region in Fig. 2(a) and 2(d), and show in Fig. 2(b) and 2(e) the temporal evolutions of the YDFL with a time window of 5000 points. Although the average powers are both about 0.28 W, the curves in the two figures are different from each other. As shown in Fig. 2(b), there exists two different kinds of pulses in the temporal evolution, i.e. pulse envelop sequence and periodic pulse sequence in one envelop. As shown in Fig. 2(e), there exists periodic pulses in the temporal evolution. The output power experiences sharp change during one cycle while keeps relatively stable between the adjacent cycles. In addition, although the adjacent pulses are similar to each other in a short time window, both the shape and peak power of the pulses become irregular in the long time window [see in Fig. 2(b)].

The power spectrum provides a direct way to obtain the potential periodic pulses in the temporal evolution. The selected frequency ranges in Fig. 2(c) and 2(f) are 5 MHz and 1 GHz, respectively. As shown in Fig. 2(c), the major frequency components of the power spectrum are around 3.3 MHz, which corresponds to a pulse period of about 0.3 μs. This quasi-periodic emission of laser pulses associated with the relaxation oscillation is the SSP phenomenon in YDFLs. As shown in Fig. 2(f), the major peaks of the frequency components are in an integral multiple of 103 MHz, which corresponds to a pulse period of about 9.68 ns. The period of the pulse is equal to one cavity round-trip time, thus this is the SML phenomenon in YDFLs.

In summary, the SSP and SML phenomena coexist in the YDFL. When the temporal evolutions are observed in different time scales, different instabilities dominate in the observation. Therefore, there exists dual time scale characteristics in the temporal evolution of the YDFL. Specifically, the temporal evolution would experience sharp change during one cavity round-trip while keep relatively stable between adjacent cavity round-trips.

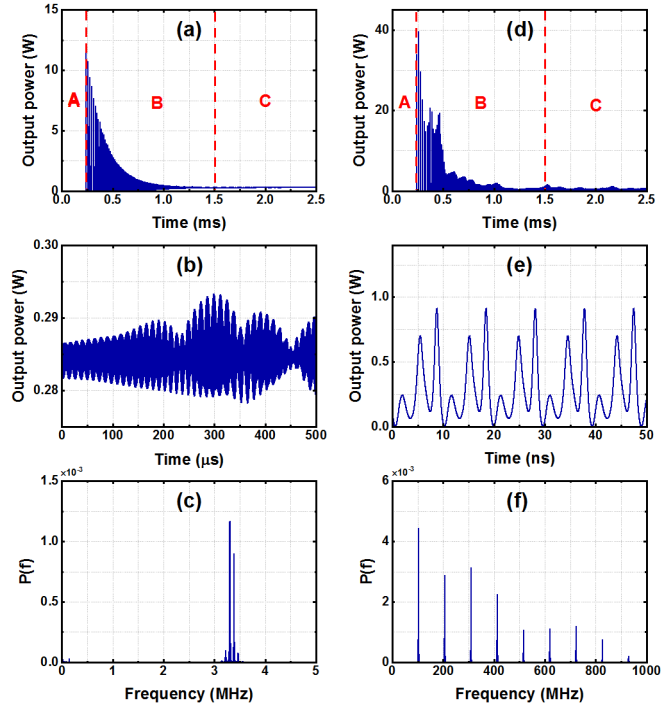

**Fig. 2.** Output characteristics in the YDFL at a pump power of 0.45 W with a time resolution of 0.1 µs (left column) and 0.01 ns (right column). The overall temporal evolution (first line), the detailed temporal evolution in relative stable region (second line), and the resulting power spectrum (third line)

Based on the temporal evolution of the optical field, the optical spectra of the YDFL could also be obtained through Fourier transform. Fig. 3 shows the corresponding optical spectral evolution of the YDFL at a pump power of 0.45 W. The simulated optical spectra are discrete and the frequency space between the spectral peaks is about 103 MHz. This value is equal to the frequency interval between two adjacent longitudinal modes. In addition, the major spectral components in the YDFL have already been formed in the relaxation oscillation region, and the output spectra keep relatively stable in the relative stable region. Thus, the longitudinal mode characteristics of the YDFL could also be included in this spatio-temporal model.

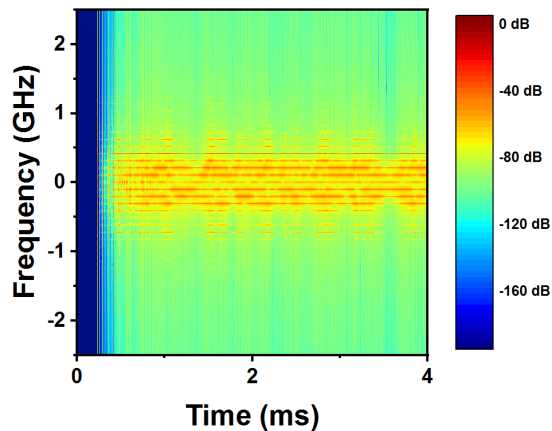

**Fig. 3.** The optical spectral evolution of the YDFL verse time at a pump power of 0.45 W

## 4. Influences of the structure parameters

The occurrences of the SML and TLP phenomena are closely related to the characteristics of the longitudinal modes in YDFLs. Accordingly, the temporal dynamics in YDFLs would change when the number of longitudinal modes increases in the output laser. Previous study has shown that the spectral width of an YDFL is nearly proportional to the bandwidth of the FBG, the nonlinear coefficient, the length of the cavity, and the output power [18]. To ensure the same frequency interval between two adjacent longitudinal modes, the structure parameters analyzed in this section are the pump power, the value of the nonlinear coefficient, and the bandwidth of the OC FBG. The relationships between the three temporal instabilities and the longitudinal mode characteristics in YDFLs would be discussed in the next section.

*4.1 Pump power*

Fig. 4 shows the average output power of the YDFL at different pump power. The average output power of the YDFL grows linearly with the pump power and the corresponding slope efficiency is about 83%.

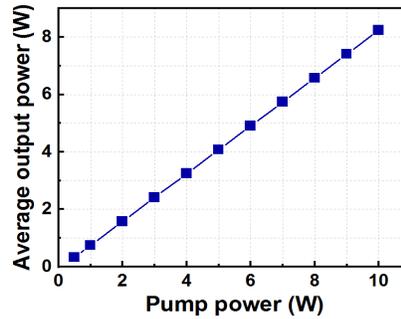

**Fig. 4.** The output power verse the pump power in the YDFL

To illustrate the possible influence of the pump power, the temporal dynamics in the YDFL are compared through the normalized temporal evolutions in the relative stable region. The output power is normalized through its maximum value here and in the following analysis.

Fig. 5 compares the normalized temporal evolutions in the YDFL at different pump powers. The left and the right columns correspond to a time resolution of 0.1 μs and 0.01 ns, respectively. When the temporal evolutions are observed with a time resolution of 0.1 μs, the SSP phenomena exist in the YDFL at different pump powers and the intensity fluctuations become stronger at higher pump powers. At a pump power over 3 W, irregular SSP pulses appear, which is similar to relaxation oscillation. Accordingly, the irregular SSP pulses might be regarded as a new relaxation oscillation in the relative stable region. When the temporal evolutions are observed with a time resolution of 0.01 ns, the SML phenomena exist in the YDFL at different pump powers. Along with the increase of the pump power, a single SML pulse gradually splits into the shorter pulses. Specifically, a single SML pulse almost completely splits into the TLP pulses at a pump power of over 5 W. Generally speaking, along with the increase of pump power, irregular SSP pulses would appear and become stronger in conventional YDFL, and a single SML pulse would gradually splits into the shorter TLP pulses.

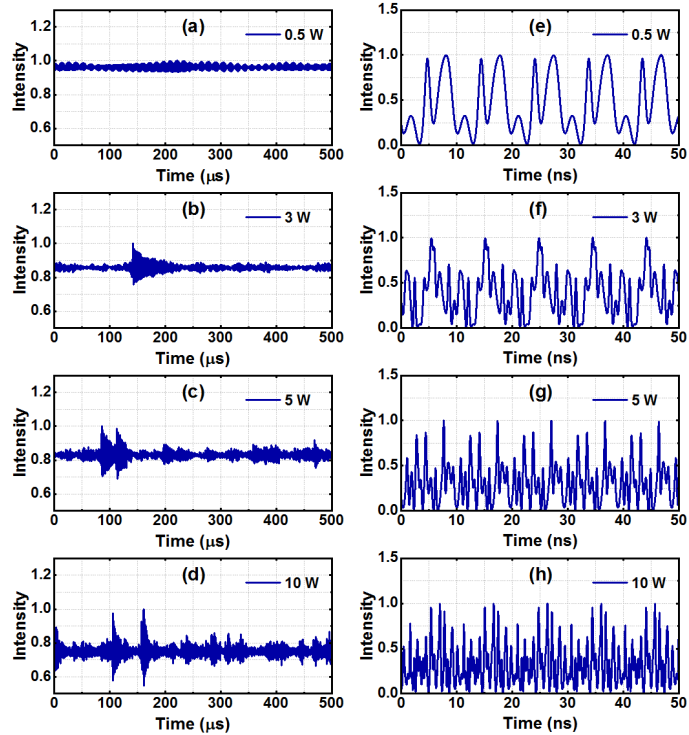

**Fig. 5.** Normalized temporal evolutions in the YDFL at different pump powers with a time resolution of 0.1 μs (left column) and 0.01 ns (right column). 0.5 W (first line), 3 W (second line), 5 W (third line), and 10 W (fourth line)

To quantify the temporal fluctuations in the YDFL, the normalized standard deviation (NSTD) is applied to compare the strength of temporal fluctuations at different pump powers. Here, the NSTD is defined as the standard deviation of the signals divided by its mean value. Fig. 6 shows the NSTDs of the temporal fluctuations in the YDFL with a time resolution of 0.1 μs (NSTD1) and 0.01 ns (NSTD2), respectively. As for NSTD1, its value stays around 0.015, begins to increase at a pump power over 3 W, and increases to about 0.04 at a pump power of 10 W. As for NSTD2, its value increases from about 0.77 (corresponds to 0.5 W) to 0.95 (corresponds to 4 W), and nearly remains the same at a pump power over 4 W.

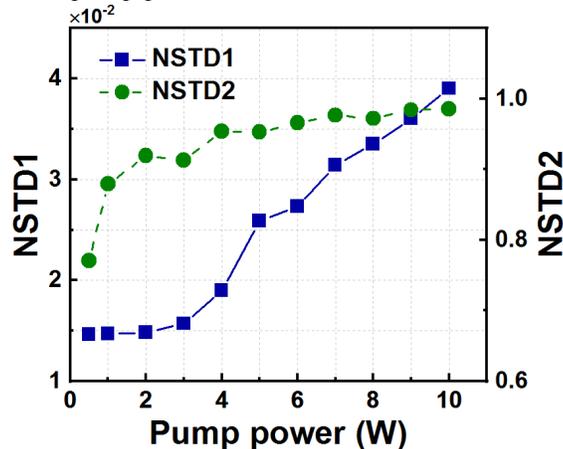

**Fig. 6.** The NSTDs of the temporal fluctuations in the YDFLs at different pump powers: NSTD 1, 0.1 μs; NSTD 2, 0.01 ns

The inflection point and the value of the NSTD1 corresponds to the occurrence and the strength of the irregular SSP pulses in the YDFL simultaneously. Thus, it is inferred that the NSTD1 is capable of quantifying the temporal fluctuations in the time scale of sub-microsecond (defined as the macroscopic fluctuations for simplicity). The value of the NSTD2 could be used to reflect the average temporal fluctuations in one cavity round trip (defined as the microscopic fluctuations for simplicity). Further, in the following section (Section V), we would show that the NSTD2 could also be used to help understanding the relationship between the temporal instabilities and the spectral properties in YDFLs.

*4.2 Value of the nonlinear coefficient*

Fig. 7 compares the normalized temporal evolutions in the YDFL at a pump power of 1 W when the values of nonlinear coefficient are set to be 0.5 $W^{-1}$/km, 3 $W^{-1}$/km, 5 $W^{-1}$/km, and 10 $W^{-1}$/km, respectively. The left and the right columns correspond to a time resolution of 0.1 μs and 0.01 ns, respectively. As for the results shown in the left column of Fig. 7, the SSP phenomena exist in the YDFL for the four cases. When the value of the nonlinear coefficient is over 3 $W^{-1}$/km, the irregular SSP pulses appear, and the intensity fluctuations in the YDFL become stronger. As for the results shown in the right column of Fig. 7, the SML phenomena exist in the YDFL for the four cases. Along with the increase of the value of the nonlinear coefficient, a single SML pulse gradually splits into the shorter pulses. Therefore, along with the increase of the nonlinear coefficient, irregular SSP pulses would also appear and become stronger in conventional YDFL, and a single SML pulse would gradually splits into the shorter TLP pulses.

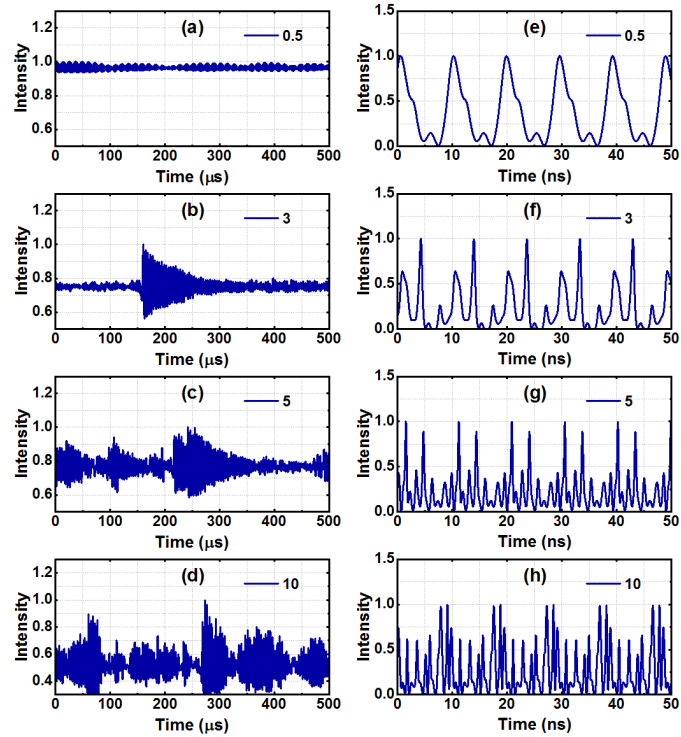

**Fig. 7.** Normalized temporal evolutions in the YDFL with a time resolution of 0.1 μs (left column) and 0.01 ns (right column) when the values of the nonlinear coefficient are: 0.5 W-1/km (first line), 3 W-1/km (second line), 5 W-1/km (third line), and 10 W-1/km (fourth line)

Fig. 8 shows the NSTDs of the temporal fluctuations in the YDFLs with different time resolutions when the values of the nonlinear coefficient are different. As for NSTD1, its value stays around 0.016, begins to increase when the value of the nonlinear coefficient is bigger than 3 $W^{-1}$/km, and increases to about 0.16 when the value of the nonlinear coefficient is 10 $W^{-1}$/km. As for NSTD2, its value increases from 0.76 (corresponds to 0.5 $W^{-1}$/km) to 0.97 (corresponds to 8 $W^{-1}$/km), and nearly remains the same when the value of the nonlinear coefficient is bigger than 8 $W^{-1}$/km.

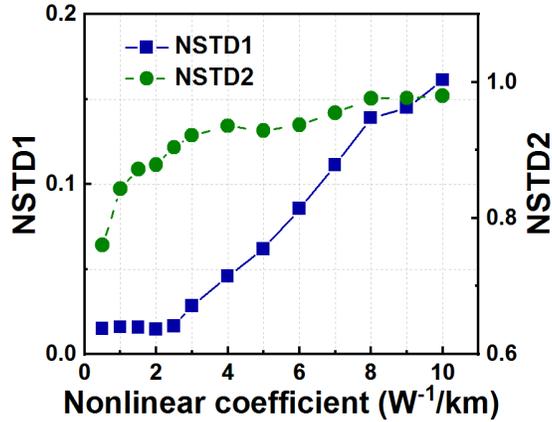

**Fig. 8.** The NSTDs of the temporal fluctuations in the YDFLs when the values of the nonlinear coefficient are different: NSTD1, 0.1 μs; NSTD2, 0.01 ns

*4.3 Bandwidth of the OC FBG*

Fig. 9 compares the normalized temporal evolutions in the YDFL when the bandwidths of the OC FBGs are 6 GHz, 10 GHz, 14 GHz, and 18 GHz, respectively. The left and the right columns correspond to a time resolution of 0.1 μs and 0.01 ns, respectively. As for the results shown in the left column of Fig. 9, the SSP phenomena and the irregular SSP pulses exist in the YDFL for the four cases. The strength of the irregular SSP pulses becomes weaker with the wider bandwidth of OC FBG. As for the results shown in the right column of Fig. 9, the SML phenomena exist in the YDFL for the four cases. Along with wider bandwidth of the OC FBG, a single SML pulse gradually splits into the shorter pulses. Therefore, along with the increase of the bandwidth of the OC FBG, a single SML pulse would gradually splits into the shorter TLP pulses, while irregular SSP pulses would become weaker in conventional YDFL.

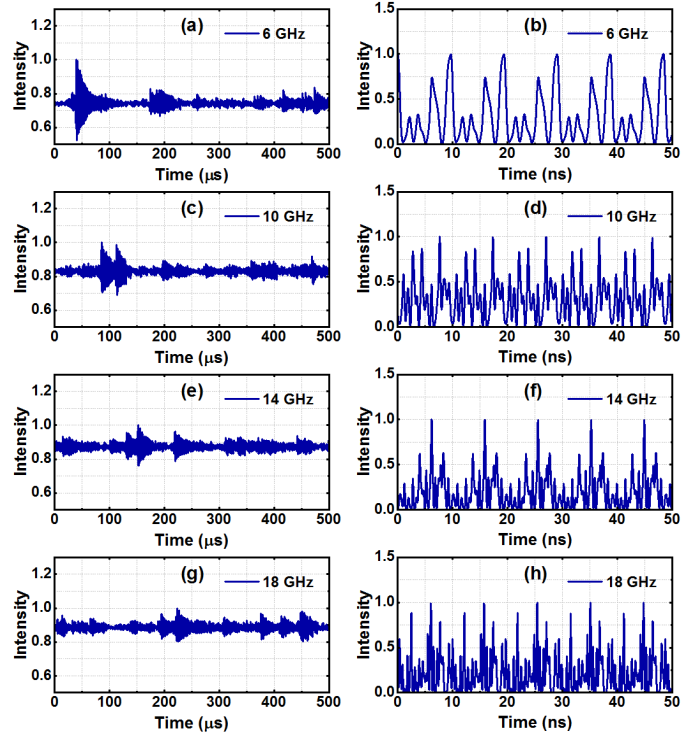

**Fig. 9.** Normalized temporal evolutions in the YDFL with a time resolution of 0.1 μs (left column) and 0.01 ns (right column) when the bandwidths of the OC FBGs are: 6 GHz (first line), 10 GHz (second line), 14 GHz (third line), and 18 GHz (fourth line)

Fig. 10 shows the NSTDs of the temporal fluctuations in the YDFLs with different time resolutions when the bandwidths of the OC FBGs are different. As for NSTD1, its overall value decreases with wider bandwidth of the OC FBG, because the strength of the irregular SSP pulses becomes weaker. As for NSTD2, its value increases slightly from 0.92 (corresponds to 6 GHz) to 0.97 (corresponds to 12 GHz), and nearly remains the same when the bandwidth of the OC FBG is over 14 GHz.

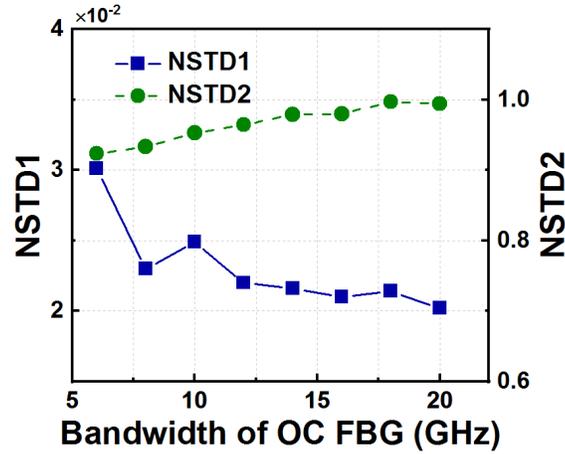

**Fig. 10.** The NSTDs of the temporal fluctuations in the YDFLs when the bandwidths of the OC FBGs are different: NSTD1, 0.1 μs; NSTD2, 0.01 ns

## 5. Relationships with the longitudinal mode characteristics

When an YDFL operates in the relative stable region, its spectral property would keep relative stable after one cavity-round trip. The temporal property and the spectral property of the optical field during one cavity-round trip satisfies the Fourier transform relationship, thus the intensities of optical field in the temporal and spectral regions are correlated to each other. Meanwhile, the temporal evolution during one cavity-round trip contains the SML and TLP phenomena in YDFLs simultaneously. Accordingly, the spectral properties of the YDFL would provide a well reference to understand the possible the relationship between the SML and TLP phenomena.

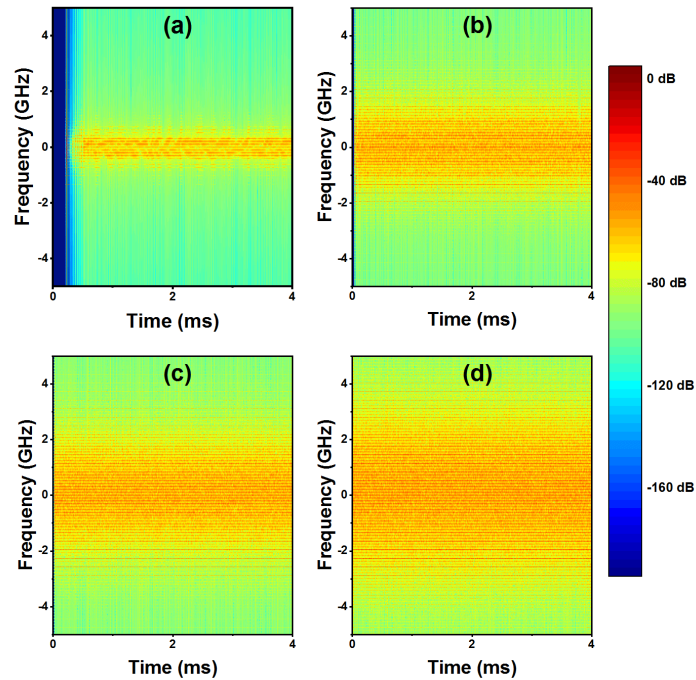

**Fig. 11.** The optical spectral evolutions of the YDFL at different pump powers: (a) 0.5 W; (b) 3 W; (c) 5 W; (d) 10 W

To do this, we first simulate the corresponding optical spectral evolutions of the YDFLs with different structure parameters. Fig. 11 shows the optical spectral evolutions of the YDFLs at different pump powers. As shown in Fig. 11, all the simulated optical spectra are discrete and the frequency spaces between the spectral peaks are equal to the frequency interval between two adjacent longitudinal modes. In addition, there exists spectral broadening phenomenon along with the increasing pump power.

The trend of the simulated spectra are all similar to the results in Fig. 5 when the values of the nonlinear coefficient or the bandwidths of the OC FBGs are different. Fig. 12 shows the root-mean-square (RMS) spectral widths of the YDFLs with different structural parameters. The square corresponds to the simulated results and the dash dot line corresponds to the linear fitting curve in Fig. 12. In all the three cases, the RMS spectral widths increase approximately linearly with the horizontal coordinates. The corresponding slopes of the linear fitting curves are 0.078 GHz/W, 0.058 GHz·W·km, and 0.069, respectively.

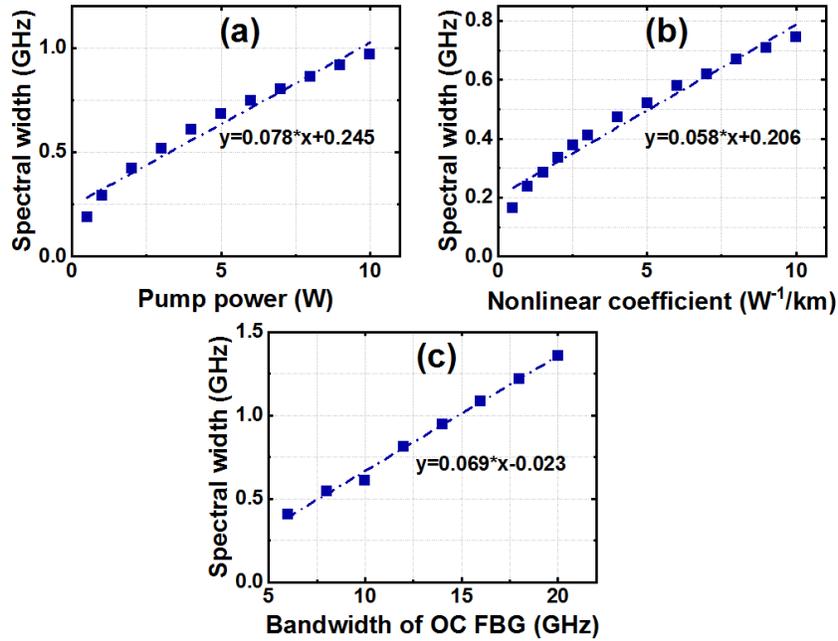

**Fig. 12.** The RMS spectral widths of the YDFLs with different structural parameters: (a) Pump power; (b) Nonlinear coefficient; (c) Bandwidth of OC FBG

When combining the NSTD2 and the RMS spectral widths in the YDFLs, the relationship between the SML and TLP phenomena with the longitudinal modes could be obtained. Fig. 13 shows the microscopic NSTDs verse the RMS spectral widths for the YDFLs with different structure parameters. As shown in Fig. 13, there exists a positive correlation between the strength of microscopic fluctuations and the RMS spectral width in YDFLs. However, when the RMS spectral width is over 0.7 GHz, it increases slightly with the RMS spectral width. The inflection point in Fig. 13 corresponds to the state when a single SML pulse almost completely splits into the TLP pulses. It distinguishes the dominant regions for the SML and TLP phenomena. In a multi-longitudinal mode YDFL, both the phases and amplitudes of the each longitudinal modes would fluctuate quickly. When the relative phase differences of some longitudinal modes keep constant, significant SML phenomenon occurs. Along with the increase number of the longitudinal modes, the relative phase differences is more difficult to keep constant, thus TLP phenomenon dominants in the observation. Consequently, the SML and TLP phenomena are the direct reflectors of the longitudinal modes superposition in multi-longitudinal mode YDFLs. Further incorporating the results in section IV, the relationship between the SSP phenomenon and the longitudinal modes could be also speculated. When the accumulated energy of the SML and TLP pulses breaks the equilibrium state of the doped ions in the fiber significantly, new relaxation oscillation reoccurs, and that is the irregular SSP phenomenon. The SSP phenomenon reflects the influence of the longitudinal modes superposition on the equilibrium state of the doped ions.

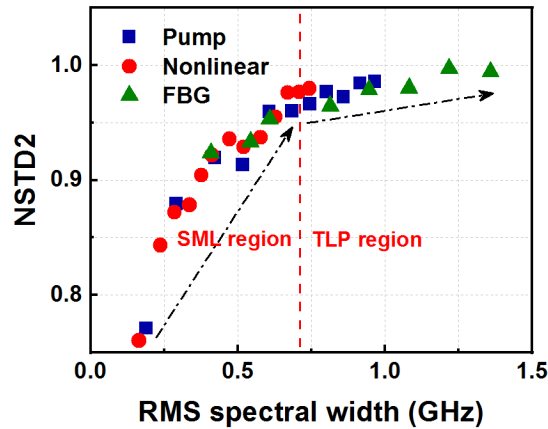

**Fig. 13.** The microscopic NSTDs verse the RMS spectral widths for the YDFLs with different structure parameters

## 6. Conclusion

In conclusion, a unified spatio-temporal model is demonstrated to tentatively unveil the temporal dynamics in multi-longitudinal mode YDFLs. Based on the theoretical analysis, it is shown that the properties of the three temporal instabilities (SSP, SML, and TLP) in the YDFLs all vary with the pump power, the value of the nonlinear coefficient, and the bandwidth of the OC FBG. The SSP phenomenon induced irregular SSP pulses mainly determines the macroscopic fluctuations of the YDFLs. The SML and TLP phenomena are closely associated to each other and dominant in different microscopic NSTD regions, which could be attributed to the number of the longitudinal modes in YDFLs. The theoretical model presented could offer a well platform for comprehensively investigate the dynamics of three typical temporal instabilities, their corresponding relationships, and possible optimizing strategies.